# BOUNDS OF CAVITATION INCEPTION IN A CREEPING FLOW BETWEEN ECCENTRIC CYLINDERS ROTATING WITH A SMALL MINIMUM GAP

A.A. Monakhov[1], V. M. Chernyavski[1], Yu. Shtemler[2]

[1]Institute of Mechanics, Moscow State University, P.O.B. B192, Moscow 119899, Russia
[2]Department of Mechanical Engineering, Ben-Gurion University of the Negev, P.O.B 653, Beer-Sheva 84105, Israel

**Abstract**

Bounds of cavitation inception are experimentally determined in a creeping flow between eccentric cylinders, the inner one being static and the outer rotating at a constant angular velocity, $\Omega$. The geometric configuration is additionally specified by a small minimum gap between cylinders, $H$, as compared with the radii of the inner and outer cylinders. For some values $H$ and $\Omega$, cavitation bubbles are observed, which are collected on the surface of the inner cylinder and equally distributed over the line parallel to its axis near the downstream minimum gap position. Cavitation occurs for the parameters $\{H,\Omega\}$ within a region bounded on the right by the cavitation inception curve that passes through the plane origin and cannot exceed the asymptotic threshold value of the minimum gap, $H_a$, in whose vicinity cavitation may occur at $H<H_a$ only for high angular rotation velocities.

## 1. Introduction

Cavitation in flows creeping through narrow passages has wide implications in industry, medicine, biology etc., and has been a focus of intensive investigation for different geometries and models. First, note the lubrication theory widely accepted up today in engineering, which employs Reynolds' conditions at the solid boundaries in order to predict cavity inception in liquids (e.g., [1], [2], [3] and references therein). On the other hand, the criterion of stress induced cavitation proposed by Winer and Bair [4] and, independently, by Joseph [5], [6] states that the flowing liquid cavitates if the principal stress of liquid exceeds the breaking strength of the liquid (the maximum tension criterion, see also recent discussions in [7], [8]). The cavitation in a creeping flow of a viscous sheared gas-saturated liquid has been observed by Kottke et al. [7] in a Couette viscometer with a thin liquid gap between the static drum and rotating inner cylinder. They suggest, according to the maximum tension criterion of cavitation inception, that the cavitation bubbles grow from preexisting wall-stabilized nuclei.

In addition to journal-bearing configurations, for which cavitation is commonly accepted to be crucial [9] cavitation induced by the rebound of a falling solid ball from a solid boundary covered with a thin layer of highly viscous liquid has also been of principal significance (e.g. [10] and [11]). In particular, Marston et al. [10] argued that the maximum tension criterion [6] provides the lower limit for the cavitation onset in their experiments. Cavitation is also found important in the dynamics of the systems of different, but qualitatively similar geometries, such as an inner sphere or cylinder free to move in creeping regime adjacent to either an inclined static plane wall coated with a thin layer of a viscous liquid, or a rotary outer horizontal cylinder filled with such liquid [12] - [21]. In particular, the role of the surface tension and inertia effects on cavitation inception in creeping flows was discussed. In all these systems, cavitation breaks the symmetry of the flow and creates a net normal component of the body weight that prevents contact between the body and underlying surface. It was found experimentally that small bubbles can appears near the downstream minimum gap position. The moving body may have quite different sizes from very small ones (e.g. [15]) to those having the same order of magnitude as the characteristic curvature radius of the underlying surface [20]. In any case, the values of angular rotation velocity and minimum gap in such systems cannot be varied independently of each other. Next, there is no quantitative data concerning the influence of the body velocity and the minimum gap on the bubble inception, and the position of cavitation bubbles within the gap between the body and the underlying surface, except some observations in Newtonian fluids, where a cavitation microbubble appears in the region of negative pressure in the near-downstream minimum gap position [13], [15], see also [19]. In qualitative correspondence with the exact Zhukovsky-Chaplygin's solution for a creeping flow between rotating eccentric cylinders generalized by [22] for the inner cylinder freely moving under the gravity force, it has been experimentally established for a sphere rolling up an inclined plane that the pressure is asymmetric with respect to the minimum gap line: it is positive in the confuser and negative in the diffuser part of the gap [13], [15]. They also propose a phenomenological flow model, according to which a bubble placed near the point of the minimum pressure does not perturb the flow, but modifies the pressure distribution and, therefore, the force acting on the sphere.

At a nanoscale level, [23] distinguish two kinds of cavitation which can occur in creeping flows through narrow gaps. They "have found that cavitation bubbles can occur either totally within the liquid, that is, away from the surfaces, or at the solid-liquid interfaces" depending on the values of "the cohesion between the liquid molecules themselves" and the "solid-liquid adhesion". At a microscopic level, the role of adhesive surface forces in cavitation inception in creeping flows through narrow passages is also well recognized (e.g. [8], [21] and references therein). In particular, two factors, the smallness of the minimum gap as compared with other characteristic lengths and the effect of the adhesive surface force, are well recognized to be crucial for cavitation inception in creeping flows. The configuration of two eccentric cylinders, the inner static one and the outer rotating at a constant angular velocity, $\Omega$, with the minimum gap, $H$, is a convenient tool to capture some principal features of cavitation inception due to the adjustability of $\Omega$ and $H$, which can both be varied as independent values during the experiments. In addition, the present study is aimed to extend our knowledge of a cavitation bubble position relative to the neighboring bubbles and within the gap.

The paper is structured as follows. In the next section, the experimental setup and main results of observations are presented. Concluding remarks are presented in Section 3.

## 2. Experimental modeling

2.1 *Experimental setup*

The experiment was performed using an outer cylinder of the inner radius $R_0 = 5 \cdot 10^{-2} m \pm 5 \cdot 10^{-5} m$, and height $5 \cdot 10^{-2} m$ made of Plexiglas, and made of aluminum an eccentric static inner cylinder of the radius $R_i = 0.375 \cdot 10^{-2} m \pm 0.2 \cdot 10^{-4} m$, and height $10^{-2} m$. The minimum gap between two vertical cylinders could be smoothly varied during the experiments, and the minimum gap decreases with the step $\sim 0.7 \cdot 10^{-5} m$. The inner cylinder was submerged into the liquid to the depth of $10^{-2} m$ from the liquid level in order to avoid end effects. The space between the cylinders was filled with degassed silicone oil polymethylsiloxane 1000 (PDMS 1000 with the kinematic viscosity $\nu = 10^{-3} m^2/s$ and the surface tension over density $\Sigma = 2 \cdot 10^{-2} m^3/s^2$ at 20°C). The present experiments were carried out for the liquid preliminary degassed by shaking the liquid volume, further foaming of the free surface, and vacuum pumping to 50$\mu Hg$. A dc motor was used to rotate the outer cylinder about its axis at the linear velocity $V_o = \Omega R_o$ varied within the range from $\Omega = 0.04$ s$^{-1}$ to $\Omega = 1$ s$^{-1}$. The bubbles dynamics, size and location in the gap between cylinders were monitored through the transparent side and top surfaces of the outer cylinder. Two video cameras were used to photograph the cavitation bubbles within the liquid-filled gap (both the side and top views A and B in Fig. 1). These video cameras allow to monitor bubbles with sizes greater than $0.2 \cdot 10^{-4} m$. The lenses were chosen with the magnification X20, two cameras were frame-synchronized, the frame frequency of cameras was 30 frames per second. The effective pixel resolution was 640x480.

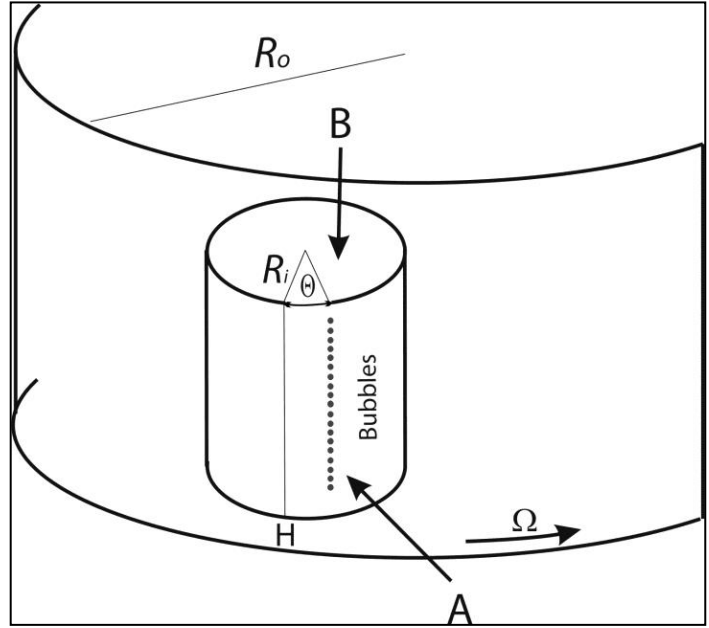

**FIG. 1.** Schematic view of the experimental setup

$\Omega$ is angular rotation velocity; $R_o$ and $R_i$ – radiuses of the outer and inner eccentric cylinders; $H$ – minimum gap between cylinders; $\theta$ – angular coordinate of the cavitation bubble center; A and B – side and top views to photograph cavitation bubbles. Solid and dotted vertical lines on the surface of the inner cylinder depict the location of minimum gap and cavitation bubbles, correspondingly.

*2.2 Experimental results for threshold of cavitation inception*

The inception of cavitation was detected by monitoring side and top views for some values of the minimum gap $H$ and angular rotation velocity, $\Omega$. Every run of the experiments with a fixed value of the angular rotation velocity starts from sufficiently large values of the minimum gap for which cavitation bubbles have not been found. Then the minimum gap is reduced up to a threshold value, at which the flow region is free from cavitation bubbles yet. After that, a small reduction in the minimum gap immediately leads to the appearance of cavitation bubbles on the surface of the inner cylinder along a line parallel to its axis. This line is located near the downstream minimum gap position. The bubbles become visible and are recorded by video cameras (see Fig. 2). Varying the rotation velocity in a certain range of values allows us to visualize the cavitation inception curve for all values of $H$ and $\Omega$ (see Fig. 3).

In Figure 2(a), a photo of four neighboring bubbles in the middle part of the inner cylinder is presented. They look as near-equally spaced approximately circular light spots of the same sizes. The distance between the neighboring bubbles' centers is found to be approximately equal to the minimum gap size, $L \approx H$, while the bubbles sizes in the vicinity of the threshold of cavitation inception are much less than $H$. If an exact threshold value of the minimum gap could be set in for any fixed value of the angular rotation velocity then the characteristic size of cavitation bubbles would be vanishingly small. A small penetration into the parameter region of cavitation leads to the appearance of cavitation bubbles registered in Fig. 2 (a). The smallest registered bubble sizes mainly depend on the effective spatial resolution of the optics.

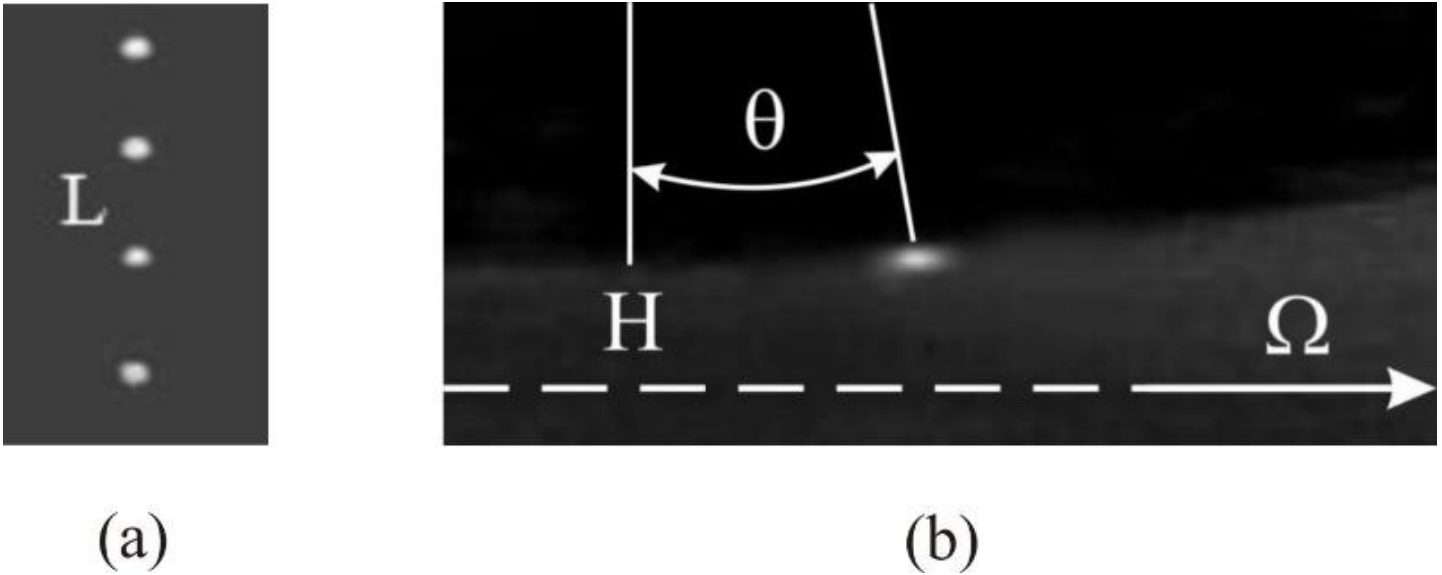

FIG. 2. Cavitation inception bubbles for the angular velocity $\Omega=0.34$ s$^{-1}$ and minimum gap $H=0.2 \cdot 10^{-3} m$.

(a) Side view: Bubbles in the mid of inner cylinder (light spots), distance between the bubbles centers,
$$L \approx H \approx 0.2 \cdot 10^{-3} m.$$

(b) Top view: Dark spot is the circular cross-section of inner cylinder; small light disc-shaped spot – cavitation bubble; horizontal dashed line – outer cylinder (schematic); $\theta \approx 7°$.

In Figure 2(b), where cavitation bubbles were photographed from the top, the solid vertical line depicts the minimum gap location. A fragment of the circular cross-section of the inner cylinder is visible as a dark spot at the top of Fig. 2(b). The wall of the outer cylinder is shown schematically by a horizontal dashed line (since the radius of the outer cylinder is significantly larger than that of the inner cylinder). The cavitation bubble located near the downstream minimum gap position is clearly visible as a light disc-shaped spot of small thickness with angular coordinates of the bubble centers that have nearly constant values within the available accuracy, $\theta \approx 7°$. Errors in the measurements of $\theta$ recorded from the top are reduced if the camera is set in front of the minimum gap line, and the location of cavitation bubble lines is registered twice at the instants of the lines formation – at the clockwise and counterclockwise rotation of the outer drum. After that, the distance between the corresponding two bubble lines, $2\Delta$, is determined and the angular coordinate of the bubble centers is calculated as follows: $\theta \approx \arcsin(\Delta/R_i)$.

In Table 1 some typical values of parameters for the cavitation inception are presented. Note that a slight variation of the minimum gap value in the vicinity of the asymptotic threshold value on the cavitation inception curve leads to a significant reduction in the corresponding threshold value of rotation velocity $\Omega$ ($\Omega=0.11 s^{-1}$ for $H=0.19 \cdot 10^{-3} m$ instead of $\Omega=0.34\ s^{-1}$ for $H=0.2 \cdot 10^{-3} m$).

TABLE 1. Typical values of parameters for cavitation inception

| $\Omega$, s$^{-1}$ | $H$, m | $\theta$, ° | $L$, m | $L/H$ |
|---|---|---|---|---|
| 0.34 | $0.20 \cdot 10^{-3}$ | 7 | $0.20 \cdot 10^{-3}$ | 1.0 |
| 0.11 | $0.19 \cdot 10^{-3}$ | 6.8 | $0.18 \cdot 10^{-3}$ | 0.95 |
| 0.03 | $0.15 \cdot 10^{-3}$ | 7.1 | $0.14 \cdot 10^{-3}$ | 0.93 |

The cavitation inception curve (see Fig. 3), which has been obtained in the plane of parameters $\{H, \Omega\}$ is bounded on the right by an asymptote $H_a = 0.21 \cdot 10^{-3} m$ ($H < H_a$), in whose vicinity a high rotation velocity is needed for the cavitation inception, while for minimum gaps larger than $H_a$ cavitation does not arise (Fig. 3).

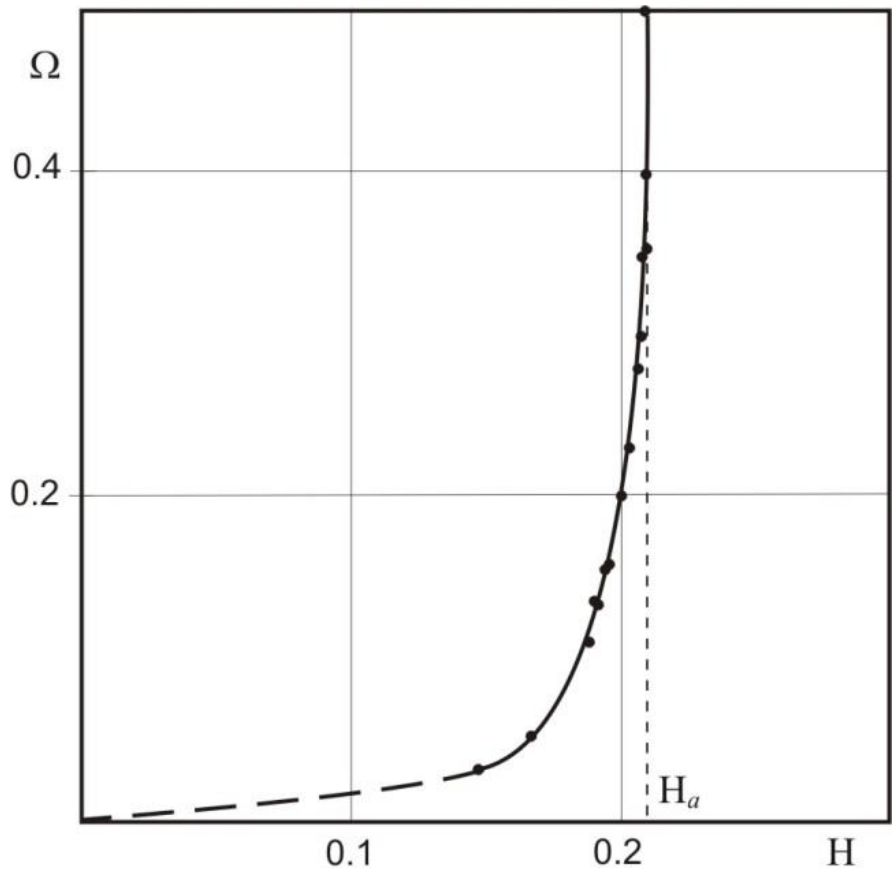

FIG. 3. Cavitation inception curve, angular roation velocity vs. minimum gap, $\Omega$ [s$^{-1}$] vs. $H$ [$10^{-3} m$]; $H_a \approx 0.21 \cdot 10^{-3} m$ is the asymptotic value of the minimum gap. Cycles are experimental threshold points for cavitation inception approximated by the best fit line passing through the origin.

The cavitation inception curve has been extrapolated to the plane origin ($H=0$, $\Omega=0$) where cavitation cannot arise. The presence of the vertical asymptote, $H=H_a$, for the cavitation inception curve in the $\{H, \Omega\}$ plane was established experimentaly. Note that with varying values of $\Omega$ and $H$ along the cavitation inception curve, the angular coordinates of the bubble centers, $\theta$, and the dimensionless ratio of the distance between the neighboring bubbles centers and the threshold values of the minimum gap, $L/H$, are nearly constant, namely, $\theta \approx 7^o \pm 0.2^o$ and $L/H \approx 1 \pm 0.1$.

## 3. Concluding remarks and discussion

Viscous-liquid flows through narrow gaps between moving solids are investigated with the angular rotation velocity and minimum gap used as adjustable parameters of the system. A short presentation of new experimental results for cavitation inception at solid-liquid-interface is given: (i) cavitation inception occurs at the surface of the inner cylinder; (ii) a cavitation inception curve is restricted from the right by a vertical asymptote in the plane of the minimum gap and angular rotation velocity.

In the present experiments with the gasified liquid, gas-vapor bubbles with a low-gas content are observed, which can be distinguished from bubbles with a significant gas-content. After a sudden stop of rotation, the low gas-content bubbles are quickly dissolved, while the significant gas-content bubbles are floated to the free surface with a reduction of their volume. It is demonstrated that cavitation can occur in a creeping flow through a gap between two eccentric cylinders, the inner one being static and the outer rotating at a constant angular velocity, $\Omega$. The geometric configuration is additionally specified by a minimum gap between cylinders small as compared with the radii of the inner and outer cylinders, i.e. $H/R_i \ll 1$ and $H/R_o \ll 1$. For some values of $H$ and $\Omega$, cavitation bubbles arise on the surface of the inner cylinder along a line parallel to its axis and are collected near the downstream minimum gap position. The threshold values for the minimum gap (separating the parameter region of cavitation from the cavitation-free one) were experimentally obtained for any given angular rotation velocity. By varying the rotation velocity, the cavitation inception curve has been obtained in the plane of $\Omega$ and $H$ bounded on the right by the asymptotic value $H_a$ – the maximal value for which the cavitation may arise at high angular rotation velocities. For all $\Omega$ and $H$ along the cavitation inception curve, the angular coordinates of the bubble centers, $\theta$, and the dimensionless ratio of the distance between the neighboring bubbles centers and the threshold values of the minimum gap, $L/H$, remain nearly constant, namely, $\theta \approx 7^o$ and $L/H \approx 1$. It is remarkable that the cavitation inception does not occur at the minimum gap position but slightly downstream. While the existence of the upper threshold, $H_a$, for the minimum gap might be anticipated, since otherwise the condition of the minimum gap smallness (discussed in Introduction) is violated, an infinitely high angular rotation velocity required for the cavitation inception in the vicinity of $H=H_a$-0 is rather unexpected.

As it was previously mentioned, an exact threshold value of the minimum gap at a fixed angular rotation velocity corresponds to vanishing cavitation bubbles. A small penetration into the parameter region of cavitation leads to the appearance of small cavitation bubbles. Regarding the observed discreteness of bubbles at the cavitation inception, one could expect that it appears as a family of equally spaced bubbles oriented along the cylinder axes. The distance between the neighboring bubbles changes with the rotation speed along the cavitation inception curve, but remains approximately equal to the minimum gap value. It can be also speculated that the near-constant distance between the cavitation bubbles could be related to the printer's instability in the system of two eccentric rotating cylinders with a small minimum gap filled with a small amount of viscous liquid (e.g. [24] and reference therein).

Let us estimate dimensionless parameters in order to specify the system configuration and flow regime. Basing on the flow similaraity in the vicinity of the minimum gap to a creeping stream through a narrow gap of the thickness $H$ between planar walls moving with a relative velocity $V_o \equiv \Omega\, R_o$, and adopting the following characteristic values (see Section 2):

$\nu = 10^{-3} m^2/s$, $\Sigma = 0.02 m^3/s^2$, $R_o = 0.05m$, $R_i = 0.00375m$, $V_o = 0.02 m/s$, $H_a = 0.2 \cdot 10^{-3} m$,

yields

$$\frac{H}{R_i} \lesssim \frac{H_a}{R_i} \sim 5.3 \cdot 10^{-2}, \qquad \frac{H}{R_o} \lesssim \frac{H_a}{R_o} \sim 4 \cdot 10^{-3}, \qquad Ca = \frac{V_o \nu}{\Sigma} \sim 10^{-3}, \qquad Re = \frac{V_o H}{\nu} \lesssim Re_a = \frac{V_o H_a}{\nu} \sim 4 \cdot 10^{-3}.$$

Thus, the minimum gap made dimensionless with the dimensional radiuses of cylinders ($H/R_i$ and $H/R_o$) as well as the Capillary and Reynolds numbers ($Ca$ and $Re$, which compare viscous stresses with interfacial ones and inertial effects with viscous ones, respectively) are small values. These values suggest that neglecting the viscous stresses and inertia effects is justified as it's widely accepted. Regarding the motivation for the choice of the silicone oil (PDMS 1000) in the present experiment, note the stability of cavitation bubbles with respect to micro-roughness of the inner-cylinder surface and to small perturbations of the minimum gap values that can be attributed to advantages of silicone oils of high viscosity. Note also that silicone fluids are commonly used for the study of cavitating systems with a narrow passage between moving surfaces ([12] - [14], [16] - [21]). In addition to our experiments with silicone oil (PDMS 1000) and an aluminum inner cylinder, several less viscous oils of 60 cSt and 100 cSt, and steel, glass, ebonite inner cylinders have been also examined. Qualitatively similar cavitation inception curves have been observed for all these systems with the only difference of characteristic scales for the axes of the minimum gap and angular rotation velocity.

Existing models of cavitation inception in the bulk of liquids are rather inapplicable to cavitation inception at the solid-liquid interface observed in our experiments, while the physical basis of such cavitation has not received sufficient attention, and the proper criterion is to be developed yet. In our view, these data require further comparative analysis, and we restrict ourselves by the presentation of typical results only with silicone oil (PDMS 1000) and an aluminum inner cylinder. We believe that experimental data presented above will support and stimulate efforts aimed at modeling cavitation inception at the solid-liquid-interface.

*Acknowledgements.* The authors are grateful to Dr. Ya. Grushevsky for his suggestions regarding the preparation of the paper. This work was partially supported by grant RFFI No. 11-08-01061-a.